\documentclass[letterpaper, 10 pt, conference]{ieeeconf}  

\IEEEoverridecommandlockouts                              
\usepackage[utf8]{inputenc}
\usepackage[T1]{fontenc}
\usepackage{float}
\usepackage{amsmath,amssymb,amsfonts}
\usepackage{algorithmic}
\usepackage{graphicx}
\usepackage{textcomp}
\usepackage{xcolor}
\usepackage{subcaption}
\usepackage{caption}
\usepackage{cite}
\usepackage{hyperref}
\def\BibTeX{{\rm B\kern-.05em{\sc i\kern-.025em b}\kern-.08em
    T\kern-.1667em\lower.7ex\hbox{E}\kern-.125emX}}

\title{\LARGE \bf 
Deep-Learning-Assisted Highly-Accurate COVID-19 Diagnosis on Lung Computed Tomography Images}
\author{Yinuo Wang, Juhyun Bae, Ka Ho Chow, Shenyang Chen and Shreyash Gupta
\thanks{All the authors are co-authors of this work with College of Engineering, Georgia Institute of Technology, Atlanta, GA 30332, USA. Email: \{ywang3781, juhyun.bae, khchow, schen801, sgupta755\}@gatech.edu. Contact ywang3781@gatech.edu for datasets and source codes.}.
}

\begin{document}

\maketitle
\thispagestyle{empty}
\pagestyle{empty}

\begin{abstract}

COVID-19 is a severe and acute viral disease that can cause symptoms consistent with pneumonia in which inflammation is caused in the alveolous regions of the lungs leading to a build-up of fluid and breathing difficulties. Thus, the diagnosis of COVID using CT scans has been effective in assisting with RT-PCR diagnosis and severity classifications. In this paper, we proposed a new data quality control pipeline to refine the quality of CT images based on GAN and sliding windows. Also, we use class-sensitive cost functions including Label Distribution Aware Loss(LDAM Loss) and Class-balanced(CB) Loss to solve the long-tail problem existing in datasets. Our model reaches more than 0.983 MCC in the benchmark test dataset. 

\end{abstract}

\section{INTRODUCTION}

Severe acute respiratory syndrome coronavirus 2 (SARS-CoV-2) infection still plays a major role in world policy changes and continues to effect billions every day. The severity of the infection rises with each day and early diagnosis can be crucial for disease control. Pandemic on such a massive scale which has impacted about 83 million people in US itself, has put primary methods of diagnosis via RT-PCR under stress. A reliable COVID-19 classification method is needed to relieve the pressure of manual clinical diagnostics. A diagnostic method of CT scans have been intensively explored in the couple of years with lung CT scans although the value of these diagnostic tests remain unknown \cite{c14}. 

\subsection{CT Scans assists in diagnosis}

COVID-19 causes symptoms consistent with pneumonia where there are inflammations caused in alveolus regions in the lungs which causes a buildup of fluid and pus leading to breathing difficulties. This buildup of pus and fluid can be detected from the chest CT scans and deep learning models can be trained to classify these scans on their severity. Based on the report by Ai et al., 2019, \cite{c15} about 60\% of the cases displayed CT features consistent with COVID-19 and almost all samples they took initial positive chest CT scan within 6 days of initial infection and 97\% of samples confirmed diagnosis from RT-PCR matched with the findings of CT scan diagnosis. Key limitation of the study remains conditions of the diagnosis at that time and the context the radiologists differentiated between viral pneumonia and COVID-19 which is something that is being considered. With the high sensitivity of diagnosing COVID-19, the report suggested that CT scans can be taken for screening, comprehensive evaluation and follow-up from the physicians. Study by Shao et al., 2020, corroboated with the findings of the Ai et al 2019 and concluded by justifying role of chest CTs as an adjunct for detection of COVID-19 infection in patients which are symptomatic with negative RT-PCR \cite{c16}.

\subsection{Literature Critique}

With the increasing number of Computed Tomography (CT) images, many reliable datasets are built, including IEEE Xplore, Web of Science, PubMed, which makes it possible to design an deep-learning based model to accurately diagnose COVID-19 based on the large-scale dataset. With increasing number of deep-learning models, it has been seen in some cases where major studies posit higher accuracy with a lack of convincing data cleaning. Several articles \cite{c17,c18} lacked a formative procedure of data-cleaning and addressing common issues handling data-imbalance and anomaly correction.

Studies have shown transfer learning models to get high sensitivity and classification between healthy, COVID-19
and pneumonia samples but fail to address the data imbalance issue \cite{c19}. Serte et al., 2021 uses similar architecture with ResNet18 and ResNet50, but selects scans based on accuracy without validation and missingness of data is not addressed \cite{c20}. STAN-CT developed in the study by Selim et al., 2020 uses generative adversarial network (GAN) to generate a standardized image which provides a better standardization and model performance but would require higher computational resources than the traditional methods \cite{c21}. Studies using the same dataset as ours (Mosmed dataset) use VGG-16 and U-net for classification got 98.11\%, 98.13\% and 97.26\% F-1 score in which our own proposed model did better with F-1 score of 99.12\% \cite{c22}. For addressing the data-imbalance issue we looked at the proposed methods of label-distribution-aware margin, focal loss and CB loss to deal at the loss level of the models in the current literature \cite{c23,c24,c25}.

The major task of the project was to design a robust end-to-end deep-learning based model and dive into some possible directions of data quality control to improve the performance of the model. Basically, the input of the model should be CT images acquired from patients, and the output of the model should be the diagnosis and severity of COVID-19 disease. Concerning that there are five kinds of ground true labels in the  IEEE  COVID-19  Hackathon  dataset (MosMedData), the model should compute the posterior probability for five classes. With the outputs of the model, the physician can make faster and more accurate decisions about the treatment. Looking at the proper lack of a standard pipeline in the current literature of CT scans, it’s necessary to develop a complete preprocessing pipeline to handle the imbalance problem and enhance the data quality, which will also shed light on the improvement of performance over other similar tasks in relevant fields.

\section{DATASET}
The dataset used to train and test our model is known as MosMedData \cite{c1}. The dataset is composed of anonymised human lung computed tomography (CT) scans with and without relevant COVID-19 symptoms. It is collected between 1st of March, 2020 and 25th of April, 2020 in different hospitals in Russia.

\begin{figure}[H] 
\centering 
\includegraphics[scale=0.5]{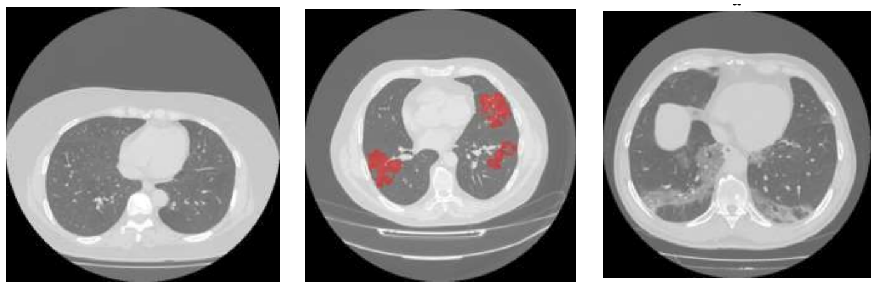}
\caption{CT scans of MosMedData} 
\label{dataset1} 
\end{figure}

The CT scans of each individual patient is stored in nii format. A given nii document contain multiple layers to describe the conditions of lungs in detail. 
Sample scans of different categories in MosMedData is shown in Fig. \ref{dataset1}.

\begin{figure}[H] 
\centering 
\includegraphics[scale=0.7]{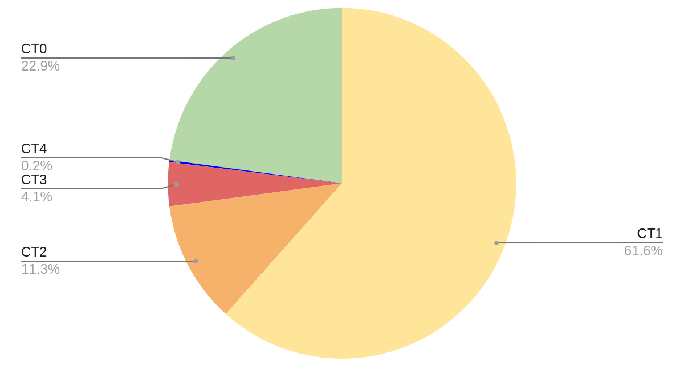}
\caption{Data distribution of MosMedData} 
\label{dataset2} 
\end{figure}

The dataset collecting 1110 patients' CT was categorized into 5 patient-level labels based on severity of CT0 (Inconsistent with pneumonia, 254 samples, 22.8\%), CT1 (Alveoli impacted in a scale of 25\% or below, 684 samples, 61.6\%), CT2 (25-50\% alveoli impacted, 125 samples, 11.3\%), CT3 (50-75\% alveoli impacted, 45 samples, 4.1\%) and CT4 (75\% and above alveoli impacted, 2 samples, 0.2\%). And each patient's CT will contain multiple scanning layers having the same label as the patient label. The data distribution is shown in Fig. \ref{dataset2}

It's clear that there exists imbalance in the dataset. CT1 takes the majority among all the categories (about 61.6 \%) while CT4 only takes a very small part in the dataset (about 0.2 \%). In this paper, considering the difficulties to conduct multi-label classification under the conditions of imbalance, we ignore the CT4 category and focus on building up an end-to-end machine learning model to classify CT0, CT1, CT2 and CT3. The techniques to handle the long-tail problem will be discussed in later sections.

\section{ARCHITECTURE(system design and baseline)}

\subsection{Workflow Design}

\begin{figure}[H] 
\centering 
\includegraphics[scale=0.26]{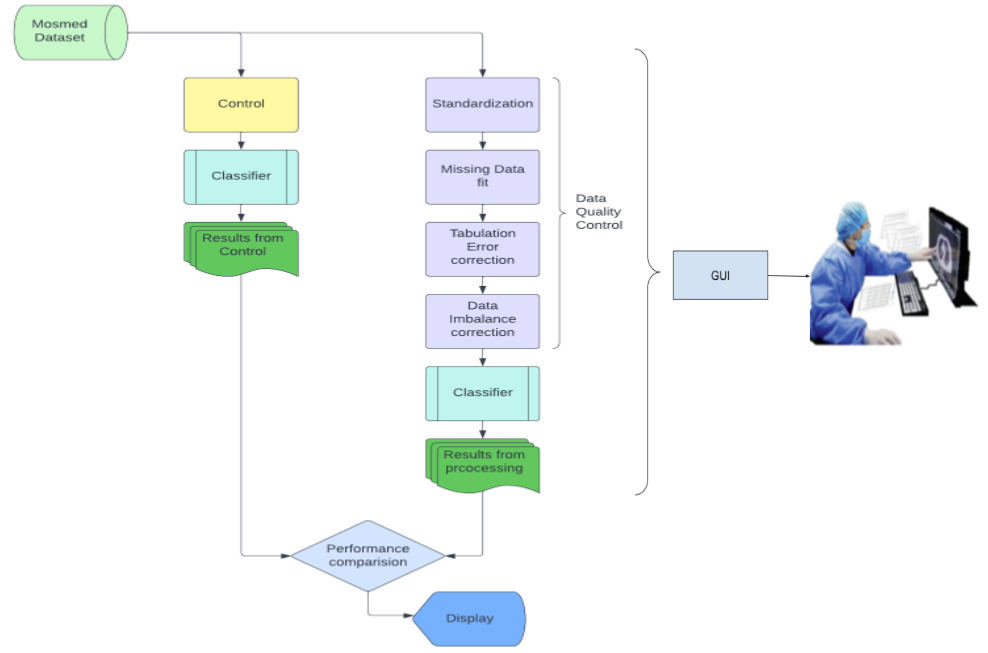}
\caption{Workflow diagram} 
\label{workflow} 
\end{figure}

The workflow diagram is described in Figure \ref{workflow}. The workflow can be divided into two sections. In the first section, we will build up a baseline model first to classify a given CT scan layer. And in the second section, before we feed the input into the baseline model, some data quality control techniques are applied to optimize the quality of input images. We will set up the same evaluation metrics for the outputs of two sections and make performance comparison accordingly. 
The input size of images is rescaled to be 224x224 to fit the requirements of backbone of model. There are three network architectures chosen as potential candidates of backbone for the baseline model: ResNet18, GoogLeNet and Vgg16. In recent years, deep learning has taken the leading role in the field of medical diagnosis. And the identity mapping of residual block can help improve the accuracy of medical diagnosis with adding more layers to the network architecture. Therefore, finally we propose a baseline model to predict different categories of CT scans based on ResNet18.

\subsection{Choice of Backbone}
ResNet was firstly proposed in 2015\cite{c2}. The network architecture is known for its simplicity and feasibility over many fields. Although many papers take ResNet50 and ResNet101, considering the light-weight characteristics and the limitation of resources, we finally chose ResNet18 as the candidate of backbone.

Driven by the motivation to handle the significance of deep layers, the residual blocks of ResNet will learn to form a residual function with reference of input x instead of learning a function without the original inputs. Under these conditions, it can allow the network architecture to contain deeper layers to boost the performance. Usually, the residual network is composed of two layers, which can be shown in [Equations \eqref{res1}]:

\begin{equation}
F(x)=W_{2 \sigma}\left(W_{1} x\right)
\label{res1}
\end{equation}

where ${\sigma}$ denotes nonlinear function ReLu. 

By adding a shortcut side, we can obtain the final output of residual block as:

\begin{equation}
y=W_{2 \sigma}\left(W_{1} x\right) + x
\label{res2}
\end{equation}

If the dimensions of input cannot fit that of output, we will use linear transformation to make adjustments:

\begin{equation}
y=W_{2 \sigma}\left(W_{1} x\right) + W_{s}x
\label{res3}
\end{equation}

\begin{figure}[H] 
\centering 
\includegraphics[width=0.5\textwidth]{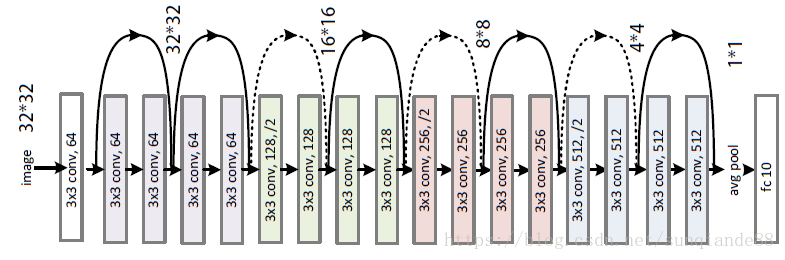}
\caption{Network Architecture of ResNet18} 
\label{network} 
\end{figure}

The network architecture of ResNet18 is described in Figure \ref{network}.

Similar to ResNet, GoogLeNet is a convolutional neural network which contains 22 layers\cite{c3}. And \cite{c3} proposes a new deep convolutional neural network architecture which is named as "Inception". The GoogLeNet can keep computational budget constant while reaching good classification performance. And another strong candidate, Vgg16, is also a deep convolutional neural network well-known for good performance over classification tasks relevant to images\cite{c4}. We conducted experiments to compare the performances of all these candidates and finally conclude that ResNet18 is the best choice for the task. More details will be discussed in later section.

\section{DATA QUALITY CONTROL}

Besides the networks we used, data quality control is also a crucial aspect which can influence the classification performance significantly. For the MosMed dataset, we mainly processed the original CT images by standardization, missing data imputation, and tabulation error fixing techniques before input the images into the networks.

\subsection{Standardization}
Although this dataset has been preprocessed to some extent by the collectors and publishers, there are still the following problems need to be solved to make the CT images become more desired and standard inputs.
\begin{itemize}

\item The original CTs contain many layers that do not belong to the lung area. For the Covid-19 disease diagnosis, only lung CTs should be used as the diagnostic basis, which means a layer selector is required to make sure only those layers of patients' lung area are used as diagnostic input. Here we implemented a range selector to filter out the first and last 25\% of the layers in CTs.  

\item The number of CT layers varies from patient to patient. Even after the CT layer selector, we cannot promise every patients have the same number of lung area CT layers as the classification input. One way can effectively unify the number of different patients' CT layers is to use polynomial interpolation on original image layers to unify CT layers into an identical number. Specifically, after this step, our method will output 10 CT slices for each patient.

\item Pixel value need to be normalized. It's a common issue for CT images because the pixel values in CT actually represent Hounsfiled Unit (HU) rather than grayscale, and the value distribution is various in different images. To achieve the value unity on the same numerical distribution while preserving the original medical information, we implement a Min-max normalization, which can be denoted as:
\begin{equation}
x^{*} = \frac{x - x_{min}}{x_{max} - x_{min}}
\label{norm}
\end{equation}
where ${x_{min}}$ and ${x_{max}}$ are the minimum and maximum value of the original CT image, and ${x}$ and ${x^{*}}$ are the original and normalized pixel value.

\item Some CTs contain abnormal image layers. Due to unstable scanning of the CT scanner or other possible reasons, some CT layers are obviously darker than others in the same patient's CT, which may send wrong information to the training network. Here, we execute a sliding window to detect such abnormal layers and using the average brightness of its adjacent layers to fix them because CT scanning is a continuous process and the adjacent layers should be similar on the image features.

\end{itemize}

\begin{figure}[h]
    \centering
    \begin{subfigure}{0.5\textwidth}
        \centering  
        \includegraphics[width=0.9\linewidth]{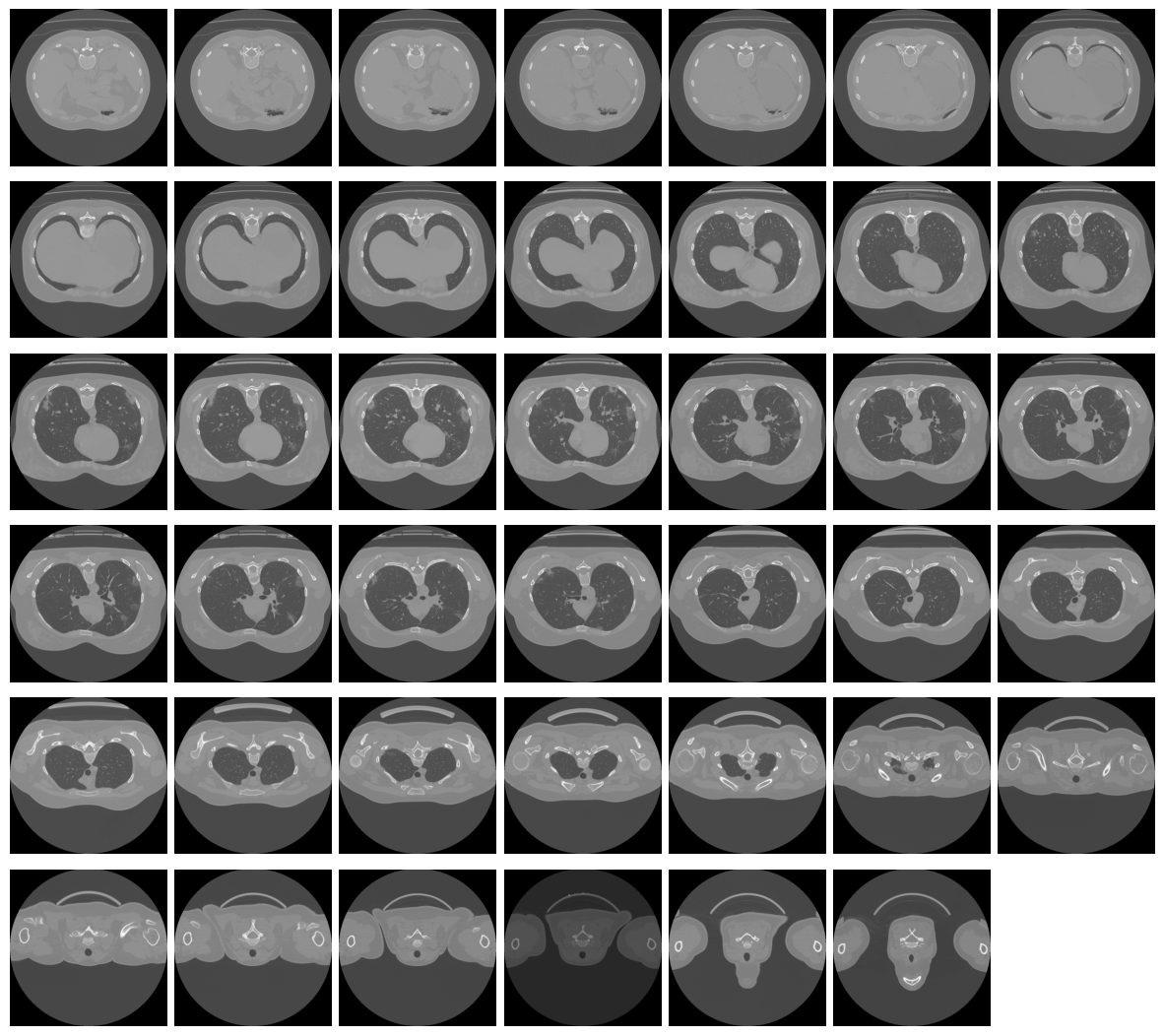}
        \caption{Original CT Images}
        \label{befstand}
    \end{subfigure}
    \begin{subfigure}{0.5\textwidth}
        \centering  
        \includegraphics[width=0.9\linewidth]{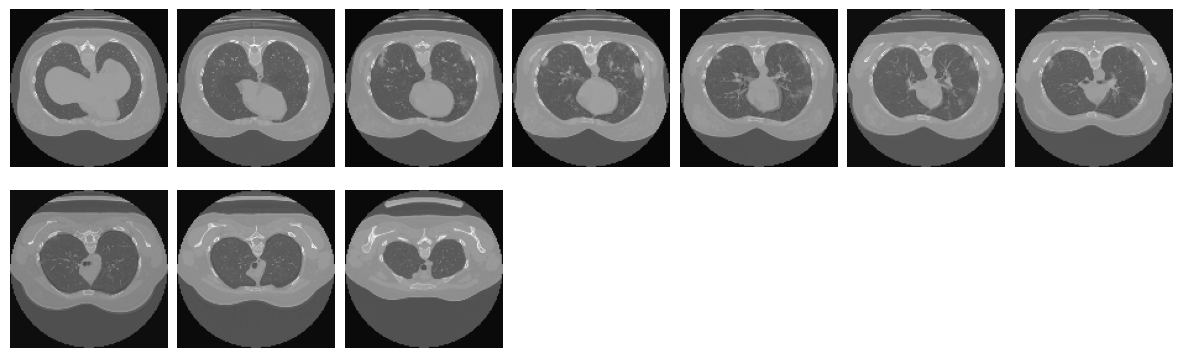}
        \caption{ Standardized CT Images}
        \label{afterstand}
    \end{subfigure} 
\caption{CT Images Standardization} 
\label{stand}
\end{figure}
After all of the above standardization processing, every patient's CT images should look like Fig. \ref{stand} b, and for comparison, the original CT images are shown in Fig. \ref{stand}a.

\subsection{Missing Data Imputation}
Learning from and predicting given incomplete data is crucial in practice because data can be missed due to various reasons and can happen in any data modality such as tabular data and image data. In the context of CTs, machine malfunctioning or even the low-dose configuration to reduce the radioactivity exposed to the patient can cause missingness in the resultant CTs. To alleviate the problem introduced by missing data, we take an imputation approach with MisGAN~\cite{c8} to enhance the quality of CTs before sending to the classifier. 

MisGAN trains three GANs: (i) synthesizing CT images, (ii) synthesizing binary masks indicating the location of a CT image where imputation needs to be conducted, and (iii) synthesizing missing details on a CT image. Similar to any GAN, the data generator $G_x$, the mask generator $G_m$, and the imputation generator $G_i$ are accompanied with their corresponding discriminator $D_x$, $D_m$, and $D_i$ respectively. 

MisGAN consists of two learning phases. With $p_D$ representing the training dataset and $p_z$ and $p_\epsilon$ denoting standard Gaussian $\mathcal{N}(0, I)$ noise distributions, it starts with training the data GAN with the loss function,
\begin{equation}
    \begin{split}
        &\mathcal{L}_x(D_x, G_x, G_m)=E_{(x, m)\in p_D}[D_x(f_\tau(x, m))]\\
        & -E_{\epsilon\sim p_\epsilon, z\sim p_z}[D_x(f_\tau(G_x(z), G_m(\epsilon)))],
    \end{split}
\end{equation}
to synthesize high-quality CT images $x$'s, where $f_\tau(x, m)=x\odot m+\tau (1-m)$, and training the mask GAN with the loss function,
\begin{equation}
    \mathcal{L}_m(D_m, G_m)=E_{(x, m)\sim p_D}[D_m(m)]-E_{\epsilon\sim p_\epsilon}[D_m(G_m(\epsilon))],
\end{equation}
to synthesize binary masks $m$'s. The two GANs are trained alternatively with the following optimization objectives:
\begin{equation}
\begin{split}
&\min_{G_z}\max_{D_x\in\mathcal{F}_z}\mathcal{L}_x(D_x, G_x, G_m),\\
&\min_{G_m}\max_{D_m\in\mathcal{F}_m}\mathcal{L}_m(D_m, G_m)+\zeta\mathcal{L}_x(D_x, G_x, G_m),
\end{split}
\end{equation}
where $\mathcal{F}_x, \mathcal{F}_m$ are defined such that $D_x$ and $D_m$ are both 1-Lipschitz in Wasserstein GANs~\cite{c9}.

After training the two GANs with $\zeta=0.2$ for $500$ epochs, the second phase is to use the two GANs trained in the first phase to teach the imputation GAN to synthesize high-quality patterns for imputation to the region of interest. It is done by formulating the corresponding loss function as follows:
\begin{equation}
    \begin{split}
        &\mathcal{L}_i(D_i, G_i, G_x)=E_{z\sim p_z}[D_i(G_x(z))]\\
        &-E_{(x, m)\sim p_D, w\sim p_w}[D_i(G_i(x, m, w))],
    \end{split}
\end{equation}
where $w$ is the uniform noise (from $0$ to $1$) introducing randomness to the imputation generator and the final optimization objectives are
\begin{equation}
\begin{split}
&\min_{G_i}\max_{D_i\in\mathcal{F}_i}\mathcal{L}_i(D_i, G_i, G_x)\\
&\min_{G_x}\max_{D_x\in\mathcal{F}_x}\mathcal{L}_x(D_x, G_x, G_m) + \phi\mathcal{L}_i(D_i, G_i, G_x),\\
&\min_{G_m}\max_{D_m\in\mathcal{F}_m}\mathcal{L}_m(D_m, G_m)+\zeta\mathcal{L}_x(D_x, G_x, G_m).
\end{split}
\end{equation}

\begin{figure}[h]
    \centering  
    \begin{subfigure}{0.5\textwidth}
        \centering 
        \includegraphics[width=0.5\linewidth]{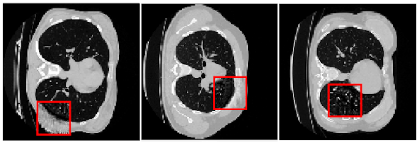}\\
        \subcaption[a]{ CT Images with Missing Data Imputation} 
        \label{missing-a}
    \end{subfigure}

    \begin{subfigure}{0.5\textwidth}
        \centering
        \includegraphics[width=0.5\linewidth]{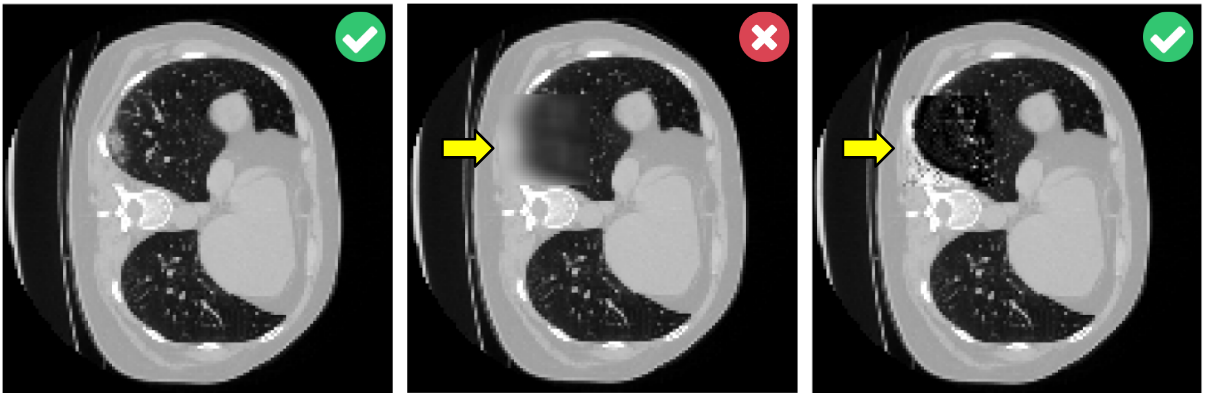}\\
        \subcaption[b]{ Original, Corrupted, and Enhanced CT Images}
        \label{missing-b}
    \end{subfigure} 
\caption{Missing Data Imputation} 
\label{missing}
\end{figure}

We train the imputation generator with $\phi=0.1$ for $500$ epochs. As shown in Fig.~\ref{missing}a where we apply a mean filter within the red block to create blurry patterns simulating the missing data scenario for visualization purposes, the imputation generator can take the observed region (i.e., outside the red block) into consideration and synthesize the imputation pattern. The enhanced CT image leads to better classification performance as shown in Fig.~\ref{missing}b where the original CT scan (left) can be correctly classified but the corrupted one (middle) is incorrectly predicted. Using the trained imputation generator, the enhanced CT image (right) can be classified again even though it was of low-quality.

\subsection{Anomaly Detection}

The detection and quantification of disease markers in imaging data is important during diagnosis, and monitoring of disease progression, or treatment response. However, data may get corrupted during the data collection and preprocessing stage. Such anomalous data may lead to incorrect machine learning training, inferences, and/or research findings. Lack of labels about the anomalousness of data points also makes researchers difficult to work on getting sufficient anomalies and correct labels are often prohibitively expensive to obtain.

GAN was originally used domains other than anomaly detection in medical imaging\cite{c10,c11,c12}. It enables to learn how to generate the realistic image. This shows the capability of GAN of capturing semantic image content and enabling vector arithmetic for generating fake images. This enables accurate detection of anomalous markers from normal anatomy. Instead of training GAN with both normal and abnormal data, this approach trains with only normal data. Then, the GAN can detect anomalousness in the image when it observes something different from the distribution for which the GAN is trained.

GAN and AnoGAN\cite{c13} are different in terms of setting. In the general GAN, when a generator creates a fake image, a discriminator tells if it is real or fake image. In AnoGAN, latent space is a vector that has the normal data feature. Therefore, the generator keeps training to generate normal data over abnormal data. If we assume that we have trained DCGAN with normal data, x is well mapped to latent space. Then, the next step is to fix the parameters in generator and discriminator, and find the optimal z value by training. 

Next, we compare real image(i.e., normal + abnormal) and fake normal image(i.e. the image from G(z1). If the fake normal image is similar to the real image, then discriminator tells normal otherwise abnormal. This is where we need to train and the authors of the paper proposed below loss function for it. It is called Residual Loss, which is the difference between the query data(i.e., manifold X of the normal real sample) and output of G(z).

\begin{equation}
    \begin{split}
        &\mathcal{L}_R(z_r)=\sum|x-G(z_r)|
    \end{split}
\end{equation}

For the next step, the paper also suggested discriminator loss function. By using it we calculate the loss in discriminator too. So, what the distriminator in AnoGAN does is that it detects how much the query image is different based on probability distribution. To do this, they define the discrimination loss as below. This is a penalty for mapping of image G(z) to the data distribution well. This extra process is called feature matching.
 
\begin{equation}
    \begin{split}
        &\mathcal{L}_D(z_r)=\sum|f(x)-f(G(z_r))|
    \end{split}
\end{equation}

For the last step, total loss is calculated by summation of Residual loss + Discriminator loss.
 
\begin{equation}
    \begin{split}
        &\mathcal{L}(z_r)=(1-\lambda)L_R(z_r) + \lambda L_D(z_r)
    \end{split}
\end{equation}
 
\begin{figure}

    \centering  
    \begin{subfigure}[b]{0.15\textwidth}
        \centering 
        \includegraphics[width=\textwidth]{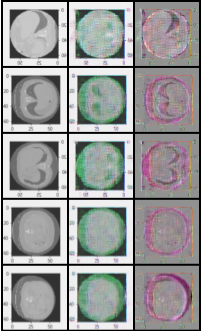}
    \end{subfigure}
    \hfill
    \begin{subfigure}[b]{0.15\textwidth}
        \centering
        \includegraphics[width=\textwidth]{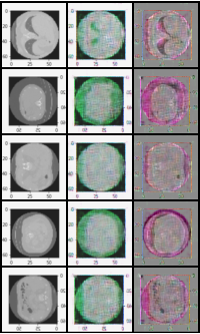}
    \end{subfigure}
    \hfill
    \begin{subfigure}[b]{0.15\textwidth}
        \centering
        \includegraphics[width=\textwidth]{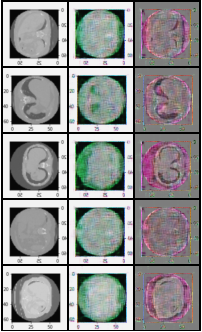}
    \end{subfigure} 
\caption{\textbf{Generated Fake Images and Anomaly Detection.} from the left original image, generated image and anomaly detection for each CT1, CT2 and CT3.} 
\label{anomaly}
\end{figure}

We trained AnoGAN with CT-0(normal) train set for 1000 epoch. As we see in Figure \ref{anomaly}, we put test image for each CT-1, CT-2 and CT-3 to the trained AnoGAN as an input. This input will be used as noise to the latent space and generate fake image to it. In the right most column we put visualization of anomaly areas based on each generated images. We also calculate anomaly score with the same test set for all classes Table \ref{anomaly_score}. Anomaly Scores for CT0, CT1 and CT2 are too similar, but CT3 can be detected by threshold of Anomaly Score.

\begin{table}[ht!]
\centering
\caption{Anomaly Score }
\begin{tabular}{ |c|c|c|c|c|}
 \hline
\bf Class &CT-0&CT-1&CT-2&CT-3\\
 \hline
\bf Average & 0.0326 & 0.0313 & 0.0308 & 0.0355 \\
 \hline
 \bf Min & 0.0217& 0.0188 & 0.0217 & 0.0205\\
 \hline
 \bf Median & 0.0282 & 0.0391 & 0.0368 & 0.0333\\
 \hline
 \bf Max & 0.0464 & 0.1157 & 0.0571 & 0.0549\\
 \hline
\end{tabular}

\label{anomaly_score}
\end{table}

\section{MEASUREMENTS TO SOLVE DATA IMBALANCE}
By analyzing the data distribution, it's clear that the CT scan resources of different categories are extremely imbalanced. The scenario will finally lead to poor generalization of decision boundaries and low accuracy of classifier. Some techniques are required to overcome the challenges of data imbalance. Basically, the relevant techniques can be summarized into three categories:i)sampling methods, including oversampling and undersampling; ii)Ensemble learning; iii)Adjusted loss functions. The first method is case-sensitive and often leads to new problems, including loss of information and information overlapping of sample space. And the second method will take great computational resources and time complexity. Therefore, we naturally introduce adjusted loss functions to overcome the long-tail problems. Here we bring three kinds of loss functions:  
Label-distribution-aware margin(LDAM) loss\cite{c5}, focal loss\cite{c6} and class-balanced loss\cite{c7}.

LDAM loss correct cross entropy loss by introducing the margin item  ${\bigtriangleup y}$, which can be denoted as:

\begin{equation}
\begin{aligned}
\bigtriangleup y_{j}=\frac{C}{n_{j}^{1 / 4}},
\end{aligned}
\end{equation}

where ${n_{j}}$ is the number of samples belonging to a given ground true category ${j}$.

The margin item makes it possible for model to generalize better. The intuition is that: Based on mathematical derivation, LDAM loss defines the relationship between margin and the number of samples belonging to a given category. The final format of LDAM loss can be described as:

\begin{equation}
\begin{aligned}
\mathcal{L}=-\log \left(\frac{e^{\frac{z_{y}-\bigtriangleup y}{s}}}{\sum_{i=1}^{C} e^{\frac{z_{i}}{s}}}\right),
\end{aligned}
\end{equation}
where ${z_{y}}$ denotes the logit score of groud true label; ${z_{i}}$ denotes the logit score of other labels, and s represents expanding factor.

Focal loss is well-known for its utility in multiple classification tasks with imbalanced datasets. It allows for decrement in weight assigned for significance of easy samples and increment in weight of significance assigned for hard samples through the training process. 

\begin{equation}
F L_{\text {loss }}=-\left(1-p_{t}\right)^{\gamma} \log \left(p_{t}\right)
\end{equation}
where gamma denotes the adjustment coefficient to adjust the weight assigned for significance, and ${p_{t}}$ denotes the prediction probability of a given category. 

The core of class-balanced loss is the definition of effective number of samples. Supposing ${E_{n}}$ denotes effective number of samples, we can compute it based on the following formula:

\begin{equation}
E_{n}=\left(1-\beta^{n}\right) /(1-\beta)
\end{equation}

where ${\beta=(N-1) / N}$.

Further we have complete form of class-balanced loss as:

\begin{equation}
\mathrm{CB}(\mathbf{p}, y)=\frac{1}{E_{n_{y}}} \mathcal{L}(\mathbf{p}, y)=\frac{1-\beta}{1-\beta^{n_{y}}} \mathcal{L}(\mathbf{p}, y)
\end{equation}

where ${n_{y}}$ is the number of samples belonging to the ground true label ${y}$. 

\begin{table}[H]
\caption{Number of samples of each category in training set and corresponding reweighting factor }
\begin{tabular}{|c|c|c|c|c|}
\hline
\textbf{}                  & CT-0   & CT-1   & CT-2   & CT-3   \\ \hline
Number of Training Samples & 2540   & 6840   & 1250   & 450    \\ \hline
Reweighting Factor    & 0.1946 & 0.1519 & 0.2324 & 0.3000 \\ \hline
\end{tabular}
\end{table}

Class-balanced loss is equivalent to introduce a reweighting vector relevant to the number of samples per class and a hyperparameter describing the effective number of samples. Accoding to \cite{c5}, reweighting is vertical to the use of margin term. Therefore the combination of these two will become possible to bring about more benefits in boosting the performance of our system.

\section{EXPERIMENTS}
\subsection{Settings of System Parameters}
To justify the parameters' influence and select most appropriate parameters for our system, we test different parameters based on grid search, especially the most crucial parameters of network, learning rate, batch size and LDAM loss function parameters. We take various combinations of one parameter represented by x-axis and another parameter represented by y-axis and test the performance based on cross validation. Here we just show the grid search results of the most sensitive parameter pair. Considering the extreme data imbalance, the bottleneck lies in the parameter choice of class-sensitive loss functions. Therefore the most sensitive parameter pair consists of the constant coefficient of margin item, which is denoted as m here, and the expanding factor s.

\begin{figure}[H] 
\centering 
\includegraphics[width=0.3\textwidth]{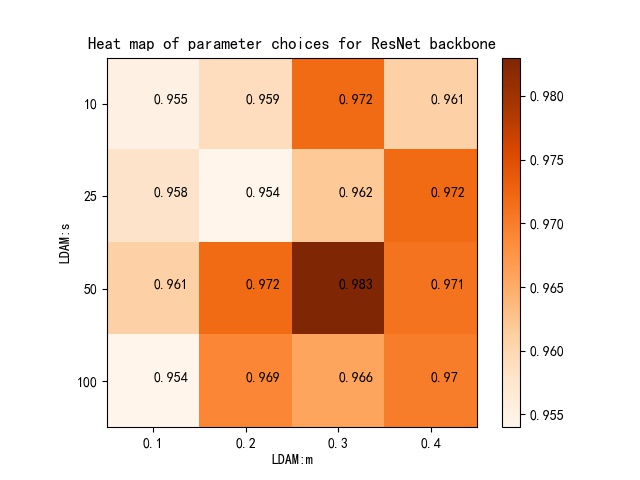}
\caption{Heat map of parameter choices for ResNet backbone} 
\label{split} 
\end{figure}

\begin{figure}[H] 
\centering 
\includegraphics[width=0.3\textwidth]{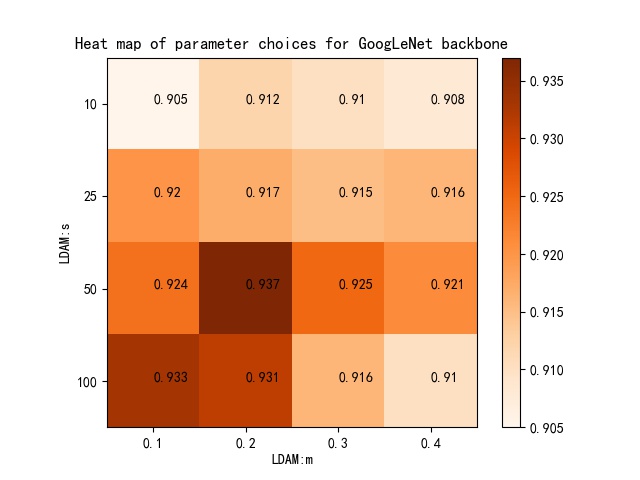}
\caption{Heat map of parameter choices for GoogLeNet backbone} 
\label{split} 
\end{figure}

\begin{figure}[H] 
\centering 
\includegraphics[width=0.3\textwidth]{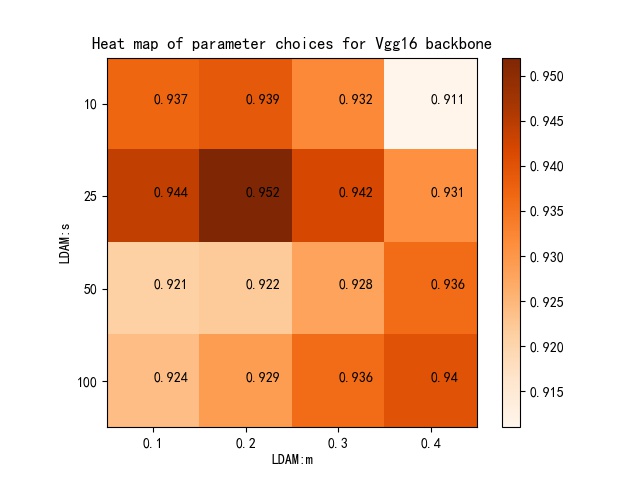}
\caption{Heat map of parameter choices for Vgg16 backbone} 
\label{split} 
\end{figure}

Based on the above experiments, the system parameters are determined and listed in Table \ref{sys}. The m here denotes the constant coefficient of margin item of label-distribution-aware loss and s denotes expanding factor\cite{c5}. Gamma is relevant to the significance of hard samples in the back propagation process when applying focal loss as the loss function of train\cite{c6}. And beta represents the capacity of sample space of datasets\cite{c7}. 

\begin{table}[H]
\centering
\caption{Settings of System Parameters}
\label{sys}
\begin{tabular}{|c|c|}
\hline
\bf Parameter      & \bf Value  \\ \hline
Learning Rate       & 0.0004 \\ \hline
Batch Size          & 32     \\ \hline
Momentum            & 0.9    \\ \hline
Weight Decay Factor & 0.001  \\ \hline
LDAM-m              & 0.3    \\ \hline
LDAM-s              & 50     \\ \hline
Gamma               & 0.02   \\ \hline
Beta                & 0.999  \\ \hline
Training Epoches    & 120    \\ \hline
\end{tabular}
\end{table}

\subsection{Evaluation}
In evaluation part, we used fourfold cross-validation for comparison of different models. To be more specific, we divided training datasets into four parts. And we selected one of the four sections in turn as the validation part. Accordingly, we can have four groups of datasets for training and validation (Fig \ref{split}).

\begin{figure}[H] 
\centering 
\includegraphics[width=0.5\textwidth]{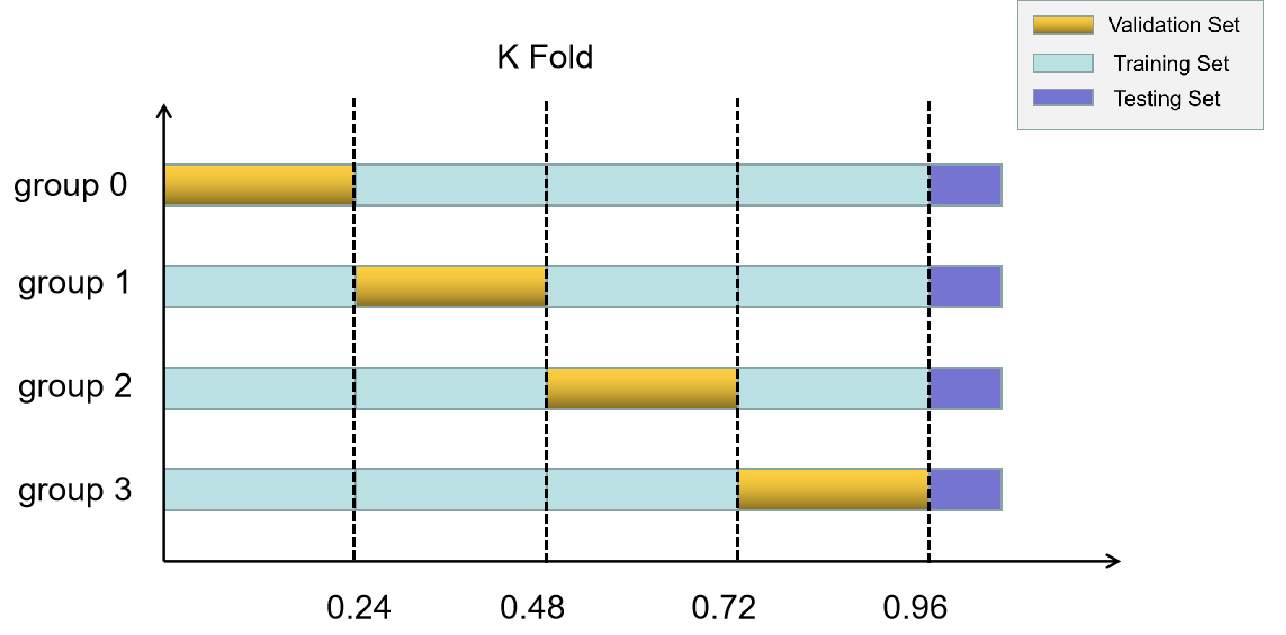}
\caption{Workflow diagram} 
\label{split} 
\end{figure}

The evaluation metrics are as following:  Matthews correlation coefficient (MCC), F1 score, balanced accuracy, precision, sensitivity and Kappa. The focus is to more accurately describe the performance over the imbalanced datesets to avoid classification bias.

True positives (TP) and true negatives (TN) are the numbers of correctly predicted samples of one specific category among CT0, CT1, CT2 and CT3 and the numbers of samples from other opposite categories, whereas false positives (FP) and false negatives (FN) are the numbers of misclassified samples of given category and other categories respectively. Based on these definitions, we can compute MCC accordingly as following:

\begin{equation}
\begin{tiny}
\begin{aligned}
\text{MMC}=\frac{\mathrm{TP} \times \mathrm{TN}-\mathrm{FP} \times \mathrm{FN}}{\sqrt{(\mathrm{TP}+\mathrm{FP}) \times(\mathrm{TP}+\mathrm{FN})\times(\mathrm{TN}+\mathrm{FP}) \times(\mathrm{TN}+\mathrm{FN})}}.
\end{aligned}
\end{tiny}
\end{equation}

Precision represents the fraction of correct classified samples over the total number of samples, and recall represents the fraction of correct predictions over the number of samples with ground true labels under a specific category. In our experiments, precision and recall score are all computed by averaging over the four categories. Also, based on precision and recall score, we can further compute ${F_{1}}$ score as:  
 
\begin{equation}
F_{1} score = \frac{2*Precision*Recall}{Precision+Recall}
\end{equation}

Kappa coefficient is computed based on two coefficients: ${p_{0}}$ and ${p_{1}}$. ${p_{0}}$ is the sum of the number of correctly classified samples of each category divided by the total number of samples, that is, the overall classification accuracy. And Supposing the number of real samples of each class is ${A_{1}}$, ${A_{2}}$, ..., ${A_{C}}$ and the predicted number of samples of each class is ${B_{1}}$, ${B_{2}}$, ..., ${B_{C}}$, we can compute ${p_{1}}$ as:

\begin{equation}
p_{0}=\frac{A_{1} \times B_{1}+A_{2} \times B_{2}+\ldots+A_{C} \times B_{C}}{n \times n}
\end{equation}

where ${n}$ denotes the total number of all the samples.

Usually we take accuracy as the evaluation metric for classification task. However with long-tail problem, it's not reasonable to follow similar evaluation methods.Instead, we use balanced accuracy to evaluate the system's performance more comprehensively, which can be denoted as:

\begin{equation}
\text { Balanced Accuracy }=\frac{\mathrm{TP}+\mathrm{TN}}{2}
\end{equation}

\subsection{Results}

Firstly, we conducted experiments over baseline models with different backbone network architectures, including ResNet18, GoogleNet and VGG16. For different models, we searched and implemented corresponding best parameters using the method we introduced in section VI-A. The results are described as following:

\begin{figure}[H] 
\centering 
\includegraphics[width=0.35\textwidth]{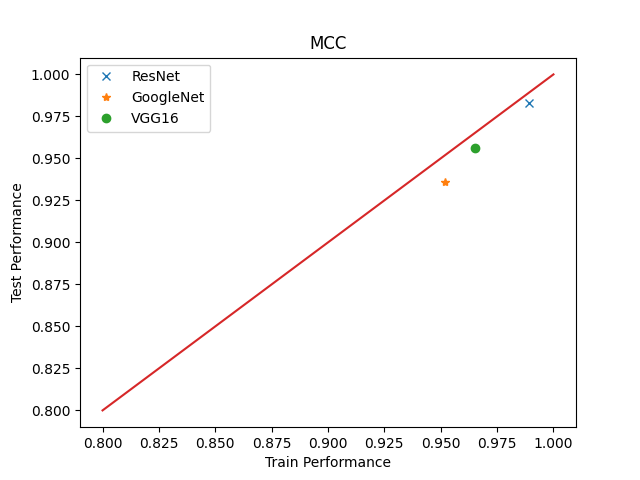}
\caption{Comparison of different models performance} 
\label{compare2} 
\end{figure}

From the figure, it's obvious that the baseline model with ResNet18 network as backbone outperforms other two. With the large datasets, shallow network is unable to build up the function to figure out the decision boundary. And ResNet makes the model more robust to solve the degeneration problems, which explains why ResNet18 can reach much better result in the task of classification of CT scans.

\begin{figure}[H] 
\centering 
\includegraphics[width=0.5\textwidth]{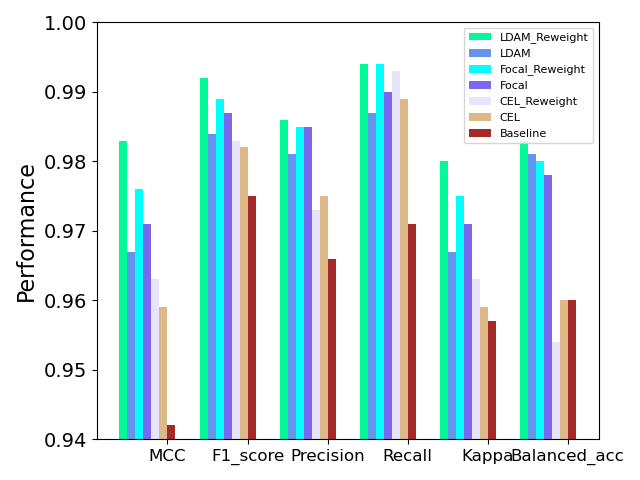}
\caption{Comparison of ResNet performance with combinations of different techniques} 
\label{compare} 
\end{figure}

And we also finished experiments to compare baseline model's performance with models processed by data quality control pipeline and trained based on adjusted loss functions. The results are shown in Figure \ref{compare}.

Other than the baseline model, all the models are processed by complete data quality control pipeline. And each model is corresponding to a specific combination of techniques to overcome the long-tail problem. The combination of LDAM loss and class-balanced reweighting vector makes the corresponding model stand out in all the evaluation metrics, which proves that the combination will be the optimal strategy when training model over imbalanced datasets. Also, baseline model's performance is much lower than all the processed models, revealing that it's meaningful and fruitful to apply appropriate data quality control measures in classification task relevant to COVID-19 CT scans.

\subsection{Diagnostic Result}
\begin{figure}[H] 
\centering 
\includegraphics[width=0.43\textwidth]{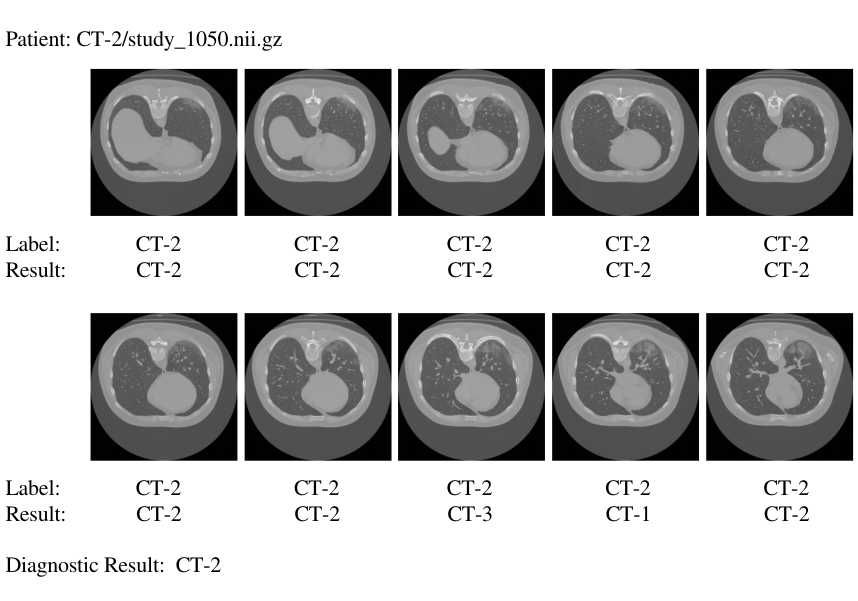}
\caption{Classification Result and Diagnostic Result} 
\label{diagnosis} 
\end{figure}

Although we have successfully trained the classifier model and test it on the random split image-level test set, the purpose of our work is to implement the model in the real world and assist doctors to make correct diagnosis. Therefore, we developed a GUI to obtain the diagnostic result on the patient-level rather than the image-level. Specifically, all the qualified and preprocessed lung CT layers will be sent in the pre-trained classifier, and the class with the most images will be output as the diagnosis result. An example is shown in Fig. \ref{diagnosis}, there are 10 preprocessed CT layers of one patient classified by the model, and the image-level classification results have been listed below each image, which can be summarized as [CT-0: 0, CT-1: 1, CT-2: 8, CT-3: 1], then this person's diagnosis will be a moderate Covid-19 patient (CT-2). 

As for the misclasssified images, one possible reason is that different areas of the patient's lungs have different levels of infection, while the label is on the patient level, which means all CT layers will have an identical label. To a certain extent, that label only indicates an average infection degree of the patient, but our network is trained as a image level classifier. However, according to the further experiments, due to our voting mechanism after image-level classification, the patient-level diagnostic result accuracy can reach 100\% though there are still some misclassified images.

\section{CONCLUSION}

In this paper, a practical deep learning based CT diagnostic system was purposed and implemented on MosMed Covid-19 dataset and several data quality control techniques were used to improve the classification performance. To optimize the performance, we compared multiple different indicators, especially MCC. After reliable evaluation and assessments, the ResNet18 was chosen as our final model because of its best comprehensive classification performance. With the support of multiple exciting techniques, our work has very encouraging potential in the real medical diagnostic area.

\bibliographystyle{IEEEtran}

\end{document}